\documentclass{iau}

\usepackage{amsmath}
\usepackage{graphicx}
\usepackage{multirow}

\newcommand{\Msun}{\rm M_{\odot}}

\begin{document}

\lefttitle{Giacomo Bortolini}
\righttitle{Unveiling IZw18 age's mystery}

\jnlPage{1}{7}
\jnlDoiYr{2025}
\doival{10.1017/xxxxx}
\volno{395}
\pubYr{2025}
\journaltitle{Stellar populations in the Milky Way and beyond}

\aopheadtitle{Proceedings of the IAU Symposium}
\editors{J. Mel\'endez,  C. Chiappini, R. Schiavon \& M. Trevisan, eds.}

\title{Unveiling IZw18 age's mystery:
Resolved Stellar Populations and Star Formation History Study with JWST/NIRCam}

\author{Giacomo Bortolini}
\affiliation{Department of Astronomy, The Oskar Klein Centre, Stockholm University, AlbaNova, SE-10691 Stockholm, Sweden}

\begin{abstract}
With its peculiar appearance, I Zw 18 has long been considered a unique example of a young galaxy in the nearby Universe. In this paper, we summarize the observational history of this famous galaxy, discuss the controversies surrounding its evolutionary state, and present new insights gained from JWST/NIRCam observations. These recent findings shed light on one of the most intriguing mysteries in extragalactic astronomy.
\end{abstract}


\maketitle

\section{Introduction}

\subsection{The curious case of I Zw 18}
\begin{figure}
    \centering
    \includegraphics[width=0.85\linewidth]{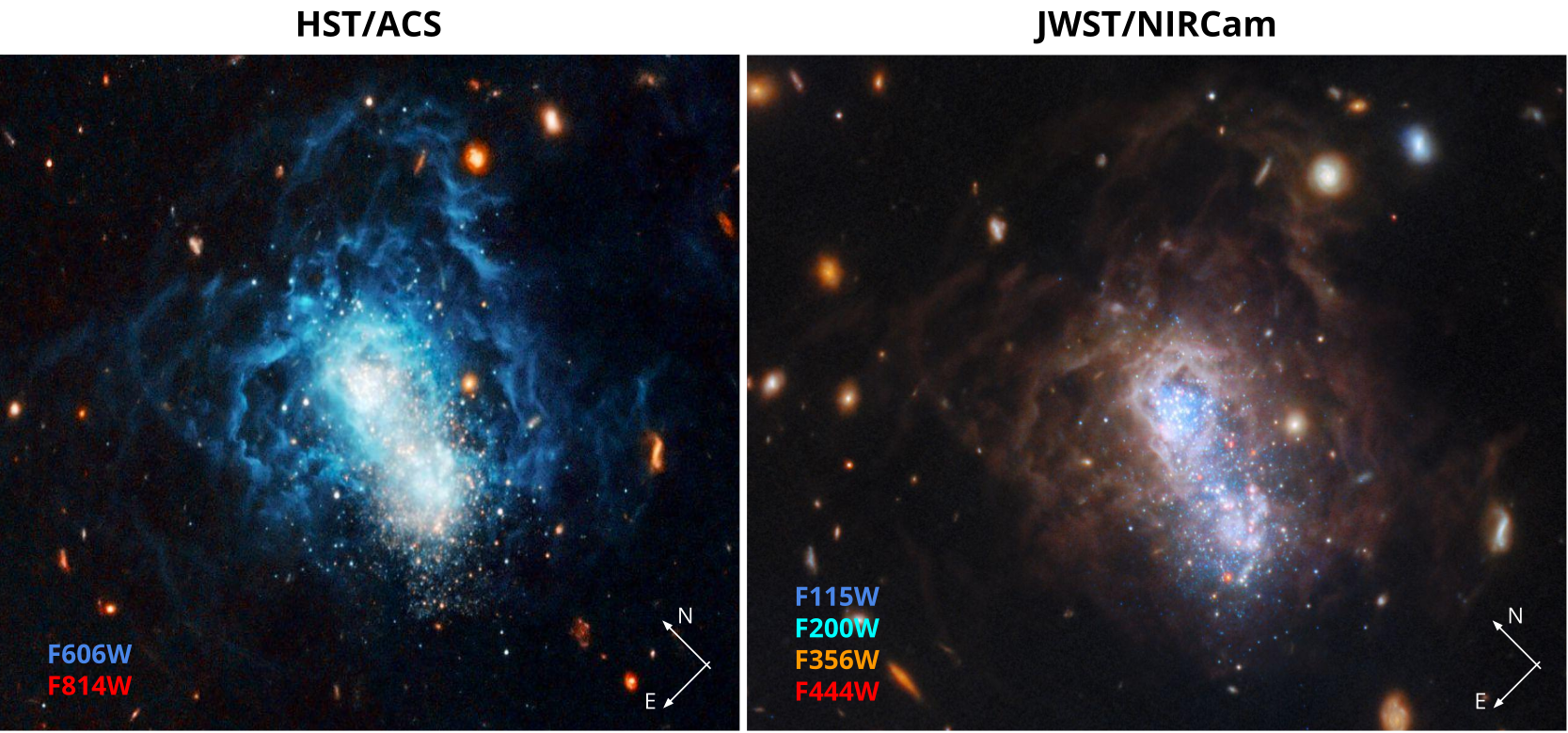}
    \caption{Comparison between color composite images of I Zw 18: the optical image from HST/ACS (left panel) and the near-infrared image from JWST/NIRCam (right panel). The filters and their associated colors are indicated in the bottom left of each panel. Image credits: ESA/NASA/CSA/STScI, \citet{Aloisi2007}, \citet{Hirschauer2024}.}
    \label{fig.1}
\end{figure}

I Zw 18, five decades after its discovery \citep{Zwicky1966}, is still the most famous blue compact dwarf (BCD) galaxy. With its extremely low oxygen abundance, i.e. ($12 + \log(O/H) \sim 7.2$ \citep{Izotov1999}, I Zw 18 is currently the third most metal-poor BCD galaxy known in the nearby Universe ($d \sim 18$ Mpc, \citealt{Aloisi2007}). I Zw 18 was originally described by \cite{Zwicky1966} as a `pair of compact galaxies'. They were later recognized as two distinct star-forming regions within the galaxy (see Figure  \ref{fig.1}), a brighter one located in the northwest (referred to as NW) and a fainter one in the southeast (referred to as SE). I Zw 18 shows blue colors ($B-V \sim -0.03$, \citealp{Vanzee1998}), high gas content, and a strikingly extended nebular emission that surrounds the two main site of star formation (see Figure \ref{fig.1}). \textsc{Hi} observations \citep{Vanzee1998,Lelli2012} have also revealed a neutral hydrogen bridge connecting I Zw 18 and a faint companion (known in the literature as Component C, \citealt{Dufour1996}), located at $\sim 22''$ in projection from the main body.

Soon after its discovery, the nature of this peculiar system sparked debates within the astronomical community. Various possible interpretations for its low metal abundance have been put forward. Some authors suggested that I Zw 18 is a genuine young galaxy caught in the process of forming its first generation of stars \citep[][among others]{Searle1972,Kunth1995,Papaderos2002,Izotov2004}. On the opposite side, several authors \citep[][among others]{Aloisi1999,Aloisi2007,ContrerasRamos2011,Annibali2013} argued that I Zw 18 may be a `mature galaxy masquerading as a toddler'\footnote{credits: NASA, ESA, and A. Aloisi.}. According to this scenario, I Zw 18 would have formed stars over a long period of time, at a rate too low to efficiently pollute the surrounding ISM \citep{Garnett1997,Legrand2001}, combined with strong stellar feedback able to removed new synthesized metals from the system \citep{Martin1996}. Regardless of the scenario, a key puzzle remains: why is this galaxy, even if a significant portion of its stellar mass might be locked up in old stars, is only recently going through a major star formation episode?

\subsection{I Zw 18's age mystery}
Since the 1990s, with the advent of the Hubble Space Telescope (HST), together with modern CCD detectors and PSF-fitting photometry routines, I Zw 18 and Component C have been resolved into their parent stars. However, I Zw 18's distance and compact nature make resolved stellar populations studies of this system rather challenging, with early photometry reaching only the more massive (i.e, younger) stars \citep{Hunter1995,Aloisi1999,Izotov2004}. However, later studies by \citet{Aloisi2007}, \citet{ContrerasRamos2011}, and \citet{Annibali2013} pushed the ACS photometry to fainter magnitudes, revealing hints of an underlying population of red and faint RGB stars in I Zw 18's color-magnitude diagram (CMD). As brilliantly expressed by \cite{Tosi2007} in the title of their paper from the IAU Symposium Proceedings No. 212, ``\textit{I Zw 18, or The Picture of Dorian Gray: The More You Watch It, the Older It Gets}''.

Today, almost 30 years after the HST launch, we can leverage the unmatched sensitivity of James Webb Space Telescope (JWST) Near Infrared Camera (NIRCam) in the near-infrared to resolve I Zw 18's old stellar populations like never before. Figure \ref{fig.1} shows the comparison between the composite color HST/ACS image \citep{Aloisi2007} in the optical (left panel) and the new composite JWST/NIRCam near-infrared image (right panel) acquired by the JWST GTO program (Program ID: 1233; PI: M. Meixner).

Here we present our study of I Zw 18's main body and Component C $1.1$ $\mu$m  $-$ $2.0$ $\mu$m  vs. $2.0$ $\mu$m CMDs, together with the analysis of the spatial distributions of their main stellar populations. Furthermore, we apply a state-of-the-art CMD fitting technique (\texttt{SFERA2.0}, \citealt{Bortolini2024b}), to reconstruct the spatially resolved star formation history (SFH) of the galaxy. Unveiling the SFH of I Zw 18 is a crucial step to understand the underlying mechanisms responsible for its extremely low metal abundance, as well as shedding light on star formation processes in extremely low metallicity environments.

\section{I Zw 18's near-infrared CMD and star formation history}
Here, we summarize the main findings and results of our work presented at the IAU Symposium 395. For a comprehensive discussion of the methodologies, analysis, and detailed results, we refer the reader to \cite{Bortolini2024}.

\begin{itemize}
    \item Both I Zw 18 and Component C near-infrared CMDs (see Figure 4 in \citealt{Bortolini2024}) are populated by stars of all ages, ranging continuously from very young upper-MS to intermediate and old age AGB and RGB stars, without any substantial gaps.
    \item Younger stars are tightly concentrated in the galaxy center, particularly tracing the NW and SE star-forming regions. Intermediate-age and old AGB and RGB stars, on the other hand, are more uniformly distributed around the center and on the outskirts of the galaxy. We observe a remarkable paucity of old stars on the NW side of the main body outskirt compared to the SE side (see Figure 5 in \citealt{Bortolini2024}). 
    \item Component C displays a mirrored stellar spatial distribution compared to the main body, with the younger stars clustered toward its South-East side.
    \item We found an overall rising trend in star formation activity in the last billion years, with two main bursts around $\sim 10$ and $\sim 100$ Myr ago, on top of what seems to be a mild but continuous activity that has lasted for the entire Hubble time with an averaged SFR of $10^{-4}-10^{-5}\, \Msun \, yr^{-1}$.
    \item I Zw 18 is now experiencing its strongest burst of star formation, located in the NW region (SFR $\sim 0.6 \, \Msun \, yr^{-1}$, in agreement with \citealt{Annibali2013} results). This burst might be driven by an interaction between the main body and Component C.
\end{itemize}

Our study confirms that I Zw 18 is not a truly young galaxy but rather an older system that has been forming stars and synthesizing metals for billions of years. This suggests that processes such as stellar winds, supernova feedback, and a possible infall of pristine \textsc{Hi} gas, likely triggered by interaction with its companion, have played a major role in removing or diluting heavy elements from the galaxy.

\end{document}